\documentclass[11pt]{article}
\usepackage[pctex32]{graphics}

\def\ben{\begin{equation}}
\def\een{\end{equation}}

\def\nn{\nonumber} \def\bd{\begin{document}} \def\ed{\end{document}}
\def\ds{\documentstyle} \let\fr=\frac \let\bl=\bigl \let\br=\bigr
\let\Br=\Bigr \let\Bl=\Bigl
\let\bm=\bibitem
\let\na=\nabla
\let\pa=\partial \let\ov=\overline
\newcommand{\be}{\begin{equation}}
\newcommand{\ee}{\end{equation}}
\def\ba{\begin{array}}
\def\ea{\end{array}}
\def\ft#1#2{{\textstyle{\frac{\scriptstyle #1}{\scriptstyle #2} } }}
\def\fft#1#2{{\frac{#1}{#2}}}
\def\del{\partial}
\def\vp{\varphi}
\def\sst#1{{\scriptscriptstyle #1}}
\def\oneone{\rlap 1\mkern4mu{\rm l}}
\def\td{\tilde}
\def\wtd{\widetilde}
\def\ie{{\it i.e.\ }}
\def\dalemb#1#2{{\vbox{\hrule height .#2pt
        \hbox{\vrule width.#2pt height#1pt \kern#1pt
                \vrule width.#2pt}
        \hrule height.#2pt}}}
\def\square{\mathord{\dalemb{6.8}{7}\hbox{\hskip1pt}}}
\newcommand{\ho}[1]{$\, ^{#1}$}
\newcommand{\hoch}[1]{$\, ^{#1}$}
\newcommand{\bea}{\setlength\arraycolsep{2pt} \begin{eqnarray}}
\newcommand{\eea}{\end{eqnarray}}
\newcommand{\ra}{\rightarrow}
\newcommand{\lra}{\longrightarrow}
\newcommand{\Lra}{\Leftrightarrow}
\newcommand{\bp}{\tilde \beta^\prime}
\newcommand{\tr}{{\rm tr} }
\newcommand{\Tr}{{\rm Tr} }
\def\0{{\sst{(0)}}}
\def\1{{\sst{(1)}}}
\def\2{{\sst{(2)}}}
\def\3{{\sst{(3)}}}
\def\4{{\sst{(4)}}}
\def\5{{\sst{(5)}}}
\def\6{{\sst{(6)}}}
\def\7{{\sst{(7)}}}
\def\8{{\sst{(8)}}}
\def\m{{\sst{(m)}}}
\def\n{{\sst{(n)}}}
\def\cA{{{\cal A}}}
\def\cB{{{\cal B}}}
\def\cF{{{\cal F}}}
\def\cG{{{\cal G}}}
\def\cH{{{\cal H}}}
\def\tV{\widetilde V}
\def\tW{\widetilde W}
\def\tH{\widetilde H}
\def\tE{\widetilde E}
\def\tF{\widetilde F}
\def\tA{\widetilde A}
\def\im{{{\rm i}}}
\def\tY{{{\wtd Y}}}
\def\ep{{\epsilon}}
\def\vep{{\varepsilon}}
\def\bD{{{\bar D}}}
\def\R{{{\mathbb R}}}
\def\C{{{\mathbb C}}}
\def\H{{{\mathbb H}}}
\def\CP{{{\mathbb C}{\mathbb P}}}
\def\RP{{{\mathbb R}{\mathbb P}}}
\def\Z{{{\mathbb Z}}}
\def\bA{{{\mathbb A}}}
\def\bB{{{\mathbb B}}}
\def\bC{{{\mathbb C}}}
\def\bD{{{\mathbb D}}}
\def\bE{{{\mathbb E}}}
\def\bZ{{{\mathbb Z}}}
\def\Re{{{\frak{Re}}}}
\def\Im{{{\frak{Im}}}}
\def\cosec{{\,\hbox{cosec}\,}}
\def\Gm{{\Gamma_{\!\! -}}}
\def\Gp{{\Gamma_{\!\! +}}}
\def\stan{{standard }}
\def\nonstan{{supernumerary }}
\def\p{{\partial}}
\def\kdel#1{{\fft{\del}{\del#1}}}

\def\bog{{Bogomolny }}
\def\om{{\omega}}

\newcommand{\nnr}{\nonumber \\}
\newcommand{\pd}{\partial}
\newcommand{\ud}{\textrm{d}}
\newcommand{\dTH}{T^{\prime \, 0}_\textrm{H}}
\newcommand{\dOi}{\Omega^{\prime \, 0}_i}
\newcommand{\bx}{{\bf x}}

\begin{document}

\vspace{5mm}
\begin{center}
{\Large \bf Entropy of black holes in the deformed
Ho\v{r}ava-Lifshitz gravity } \vspace{12mm}

{\large   Yun Soo Myung \footnote{e-mail
 address: ysmyung@inje.ac.kr}}
 \\
\vspace{10mm} {\em Institute of Basic Science and School of
Computer Aided Science \\ Inje University, Gimhae 621-749, Korea}
\end{center}

\begin{center}

\underline{Abstract}
\end{center}

  We find the entropy of Kehagias-Sfetsos black hole in the deformed Ho\v{r}ava-Lifshitz  gravity by using
  the first law of thermodynamics. When applying  generalized uncertainty principle (GUP) to Schwarzschild black hole,
  the entropy $S=A/4+(\pi/\omega)\ln(A/4)$  may be interpreted as the
GUP-inspired black hole entropy. Hence, it  implies that the duality
in the entropy between the Kehagias-Sfetsos black hole  and
GUP-inspired Schwarzschild black hole is present.

\vspace{15pt}

\thispagestyle{empty}





\newpage
\section{Introduction}
Recently Ho\v{r}ava has proposed a renormalizable theory of
gravity at a Lifshitz point~\cite{ho1,ho2},  which  may be
regarded as a UV complete candidate for general relativity. At
short distances the theory of  Ho\v{r}ava-Lifshitz (HL) gravity
describes interacting nonrelativistic gravitons and is supposed to
be power counting renormalizable in (1+3) dimensions. Recently,
its black hole solutions has been intensively investigated
~\cite{LMP,CCO1,CLS,CY,MK,CCO2,Gho,KS,Myung,CJ1,CJ2,CW,park,BGS,Gho09,CL,PW,lkme}.

Concerning the spherically symmetric solutions, L\"u-Mei-Pope (LMP)
have obtained the black hole solution with dynamical parameter
$\lambda$ in asymptotically Lifshitz spacetimes~\cite{LMP} and
topological black holes were found in \cite{CCO1}. Its
thermodynamics were studied in \cite{MK,CCO2} but there remain
unclear issues in defining the ADM mass and entropy. On the other
hand, Kehagias and Sfetsos (KS)  have found the ``$\lambda=1$" black
hole solution in asymptotically flat spacetimes using the deformed
HL gravity with parameter $\omega$~\cite{KS}. Its thermodynamics was
 defined in Ref.\cite{Myung}. Also, Park has obtained  the
$\lambda=1$ black hole solution with two parameters $\omega$ and
$\Lambda_W$~\cite{park}.

It seems that the GUP-inspired  Schwarzschild black hole was related
to black holes in the deformed HL gravity~\cite{Myung}. We could not
confirm a solid connection between the GUP~\cite{gup1,gup2} and the
black hole of deformed HL gravity, although partial connections were
established between them. However,  it was known that the GUP
provides naturally a UV cutoff to the local quantum field theory as
quantum gravity effects~\cite{CMOK,KLM2}. Also,  the GUP density
function may be replaced by a cutoff function for the
renormalization group study of deformed HL gravity~\cite{myunggup}.
We have found GUP-inspired graviton propagators and compared these
with UV-tensor propagators in the deformed HL gravity. Two were
similar, but the $p^5$-term arisen from Cotton tensor was missed in
the GUP-inspired graviton propagator. This shows that a
power-counting renormalizable theory of the HL gravity is closely
related to the GUP.

In this Letter, we will make a further progress on exploring the
connection between the GUP and  black hole of the deformed HL
gravity. We obtain  the entropy of KS  black hole in the deformed HL
gravity. This entropy  may  be interpreted as the GUP-inspired black
hole entropy   when
  applying the GUP to Schwarzschild black hole.

 \section{HL gravity}
Introducing the ADM formalism where the metric is parameterized
\be ds_{ADM}^2= - N^2  dt^2 + g_{ij} \Big(dx^i - N^i dt\Big)
\Big(dx^j - N^j dt\Big)\,, \ee
the Einstein-Hilbert action can be expressed as
\be \label{Eins} S_{EH} = \fft{1}{16\pi G} \int d^4x \sqrt{g} N
\Big[K_{ij} K^{ij} - K^2 + R - 2\Lambda\Big], \ee
where $G$ is Newton's constant and extrinsic curvature $K_{ij}$
takes the form
\be K_{ij} = \fft{1}{2N} \Big(\dot g_{ij} - \nabla_i N_j -
\nabla_j N_i\Big)\,. \ee
Here, a dot denotes a derivative with respect to $t$. An action of
the non-relativistic renormalizable gravitational theory  is given
by~\cite{ho1} \be S_{HL}=\int dtd^3x \Big[{\cal L}_K + {\cal
L}_V\Big],  \label{action1} \ee where the kinetic terms are given
by \be {\cal L}_K =\frac{2}{\kappa^2}\sqrt{g} N K_{ij}{\cal
G}^{ijkl}K_{kl}= \frac{2}{\kappa^2}\sqrt{g}
N\Big(K_{ij}K^{ij}-\lambda K^2\Big), \ee with the DeWitt metric
 \be {\cal G}^{ijkl}=
\frac{1}{2}\Big(g^{ik}g^{jl}-g^{il}g^{jk}\Big)-\lambda
g^{ij}g^{kl} \ee
 and its inverse metric
 \be {\cal
G}_{ijkl}=\frac{1}{2}\Big(g_{ik}g_{jl}-g_{il}g_{jk}\Big)-\frac{\lambda}{3\lambda-1}g_{ij}g_{kl}.\ee

The potential terms is determined by the detailed balance
condition  as \bea {\cal L}_V=-\frac{\kappa^2}{2}\sqrt{g}N
E^{ij}{\cal G}_{ijkl}E^{kl}&=&
\sqrt{g}N\Bigg\{\frac{\kappa^2\mu^2}{8(1-3\lambda)}\Big(\frac{1-4\lambda}{4}R^2
+\Lambda_WR-3\Lambda_W^2\Big)\nn \\
 &-&\frac{\kappa^2}{2w^4} \left(C_{ij}
-\frac{\mu w^2}{2}R_{ij}\right) \left(C^{ij} -\frac{\mu
w^2}{2}R^{ij}\right) \Bigg\}.\label{action2} \eea Here the $E$
tensor is defined by \be E^{ij}=\frac{1}{w^2} C^{ij}-\frac{\mu}{2}
\Big(R^{ij}-\frac{R}{2} g^{ij}+\Lambda_Wg^{ij}\Big) \ee with the
Cotton tensor $C_{ij}$ \be
C^{ij}=\frac{\epsilon^{ik\ell}}{\sqrt{g}}\nabla_k\left(R^j{}_\ell
-\frac14R\delta_\ell^j\right).\label{def.K.C} \ee  Explicitly,
$E_{ij}$ could be derived  from the Euclidean topologically
massive gravity \be E^{ij}=\frac{1}{\sqrt{g}} \frac{\delta
W_{TMG}}{\delta g_{ij}} \ee with \be W_{TMG}=\frac{1}{w^2} \int
d^3 x \epsilon^{ikl}\Big(\Gamma^m_{il}\partial_j
\Gamma^l_{km}+\frac{2}{3} \Gamma^n_{il} \Gamma^l_{jm}
\Gamma^m_{kn} \Big)- \mu \int d^3x \sqrt{g}(R-2\Lambda_W), \ee
where $\epsilon^{ikl}$ is a tensor density with
$\epsilon^{123}=1$.

In the IR limit,  comparing ${\cal L}_0$ with Eq.(\ref{Eins}) of
general relativity, the speed of light, Newton's constant and the
cosmological constant are given by
\be c=\fft{\kappa^2\mu}{4}
\sqrt{\fft{\Lambda_W}{1-3\lambda}}\,,\qquad
G=\fft{\kappa^2}{32\pi\,c}\,,\qquad \Lambda_{\rm cc}=\ft32
\Lambda_W\,.\label{cg} \ee The equations of motion were derived in
\cite{cos1} and \cite{LMP}. We would like to mention that the IR
vacuum of this theory is Lifshitz  spacetimes~\cite{MK}. Hence, it
is interesting to take a limit of the theory, which may lead to  a
Minkowski vacuum in the IR sector. To this end, one may deform the
theory by introducing a soft violation term of ``$\mu^4R$"
$(\tilde{{\cal L}}_V={\cal L}_V+\sqrt{g}N \mu^4R)$ and then, take
the $\Lambda_W \to 0$ limit~\cite{KS}. We call this as the
``deformed HL gravity". This theory  does not alter the UV
properties of the HL gravity, while it changes the IR properties.
That is, there exists a Minkowski vacuum, instead of Lifshitz
vacuum. In the IR limit, the speed of light and Newton's constant
are given by
\be c^2=\fft{\kappa^2\mu^4}{2},~ G=\fft{\kappa^2}{32\pi\,c},
~\lambda=1.\label{kh} \ee

\section{Entropy of KS black hole}
A spherically symmetric solution to the deformed HL gravity was
obtained by considering the line element
\begin{equation}
ds^2 = -N(r)^2 dt^2 + \frac{dr^2}{f(r)} + r^2 \left ( d \theta^2 +
\sin^2 \theta d \phi^2 \right ) . \label{sph_ansatz}
\end{equation}
In this case, we have $K_{ij}=0$ and $C_{ij}=0$. Hence, it is
emphasized  that  we have relaxed both the projectability
restriction and detailed balance condition~\cite{ho1,muko} since
 the lapse function $N$  depends on the spatial
coordinate $r$ as well as a soft violation term of $\mu^4 R$ is
included. Substituting the metric ansatz (\ref{sph_ansatz}) into
$\tilde{{\cal L}}_V$ with ${\cal L}_K=0$, one has the reduced
Lagrangian \be \label{react} \tilde{{\cal L}}_V=\frac{ \mu^4 N
}{\sqrt{f}}\Bigg[ \frac{\lambda-1}{2\omega_\lambda} f'^2 -
\frac{2\lambda (f-1)}{\omega_\lambda r}f'+
\frac{(2\lambda-1)(f-1)^2}{\omega_\lambda r^2}-2(1 - f - r f')\Bigg]
\ee where a parameter $\omega_\lambda=8\mu^2(3\lambda-1)/\kappa^2$
specifies the deformed HL gravity.

For $\lambda=1$,  the KS  solution is given by~\cite{KS}
\begin{equation}
f_{KS} =N^2_{KS}= 1 + \omega r^2 \left ( 1 - \sqrt{1 + \frac{4
M}{\omega r^3}}\right ) \label{sph_sol}
\end{equation}
with $\omega(=\omega_{\lambda=1}) = 16 \mu^2 / \kappa^2$. In the
limit of $\omega \to \infty$ (equivalently, $\kappa^2 \to 0$), it
reduces to  the Schwarzschild metric function \be
 f_{Sch}(r)=1-\frac{2M}{r}. \ee
  From the condition of $f_{KS}(r_\pm)=0$, the
outer (inner) horizons are given by \be \label{sol}
r_\pm=M\Bigg[1\pm \sqrt{1-\frac{1}{2\omega M^2}}\Bigg]. \ee In
order to have a black hole solution, it requires that \be M^2 \ge
\frac{1}{2\omega}. \ee
 Furthermore, the  extremal black hole
 is obtained from the condition of degenerate horizon $(
 f_{KS}(r_e)=0,~f_{KS}'(r_e)=0)$ as
 \be
 r_e=M_e=\frac{1}{\sqrt{2\omega}}\ee
with $f''_{KS}(r_e)=4\omega/3$.

Thermodynamic quantities of  mass, temperature, and heat capacity
for the KS black hole are defined  as~\cite{Myung} \be
M(r_\pm)=\frac{1+2\omega r_\pm^2}{4\omega r_\pm},~~T=\frac{2\omega
r_+^2-1}{8\pi r_+(\omega
r_+^2+1)},~~C=-\frac{2\pi}{\omega}\frac{(\omega r_+^2+1)^2(2\omega
r_+^2-1)}{2\omega^2 r_+^4-5\omega r_+^2-1}. \ee In the limit of
$\omega \to \infty$, these reduce to  corresponding quantities of
Schwarzschild black hole as \be M\to \frac{r_+}{2},~~T\to
\frac{1}{4\pi r_+},~C\to -2\pi r_+^2. \ee

Now we wish to derive the entropy by considering that  the first law
of thermodynamics holds for  black hole in the deformed HL gravity:
\be dM=TdS.\ee Then, the entropy is calculated as \be S=\int dr_+
\Bigg[\frac{1}{T}\frac{\partial M}{\partial r_+}\Bigg]+S_0,\ee which
leads to~\cite{CL} \be S=\pi
\Big[r_+^2+\frac{1}{\omega}\ln\Big(r_+^2\Big)\Big]+S_0.\ee If one
chooses \be S_0=\frac{\pi}{\omega} \ln \pi,\ee then we have a
compact expression of the entropy \be \label{entropy}
S=\frac{A}{4}+\frac{\pi}{\omega}\ln\Big[\frac{A}{4}\Big]\ee with
$A/4=\pi r_+^2 $ and $G=1$. We note that in the limit of $\omega \to
\infty$, Eq. (\ref{entropy}) reduces to the Bekenstein-Hawking
entropy of Schwarzschild black hole as \be S_{BH}=\frac{A}{4}. \ee
It is clear that the logarithmic term represents the feature of KS
black hole in the deformed HL gravity. Accordingly, we have to
interpret this logarithmic term to understand why the entropy of KS
black hole takes the form (\ref{entropy}).

\section{GUP-inspired  Schwarzschild black hole}

A meaningful prediction of various theories of quantum gravity
(string theory and loop quantum gravity) and black holes is the
presence of a minimum measurable length or a maximum observable
momentum. This has provided the GUP  which modifies commutation
relations between position coordinates and momenta. Also  the
black hole solution of deformed  HL gravity reminds us the
Schwarzschild black hole modified with the  GUP~\cite{Myung}.
Hence, it is very interesting to develop
 a  close connection between GUP and
HL gravity.  A generalized commutation relation\footnote{
 We note that the GUP is in the heart of the
quantum gravity phenomenology. Certain effects of quantum gravity
are universal and thus, influence almost any system with a
well-defined Hamiltonian~\cite{DV}. In general, the GUP satisfies
the modified Heisenberg algebra~\cite{CMOK,KLM2}
 \bea
&&[x_i,p_j]=i\hbar \Big(\delta_{ij}+\beta p^2 \delta_{ij}+\beta' p_i p_j\Big),\nn \\
\label{UVCR}&&[x_i,x_j]=i\hbar \frac{(2\beta-\beta')+(2\beta+\beta')\beta p^2}{1+\beta p^2} \Big(p_i x_j-p_j x_i\Big),  \\
&&[p_i,p_j]=0, \nn \eea where $p_i$ is considered as the momentum at
high energies and thus, (\ref{UVCR}) can be interpreted to be the
UV-commutation relations.  In order to achieve the commutativity, we
have to choose $\beta'=2\beta$. In this case, the minimal length
 which follows from the modified Heisenberg algebra is given by
$
 (\Delta x)_{\rm min}=\hbar \sqrt{5\beta}$.
 We emphasize that {\it the presence of the minimal length represents  a feature of the GUP}.
 In order to see the GUP-inspired black holes, it is sufficient to consider $\beta p^2$
 because this term determines the minimal size of the black hole.} of
         \begin{equation} \label{crgup}
         [x_i, p_j]=i\hbar \delta_{ij}\Big(1+\beta p^2\Big)
         \end{equation}
         leads to the  generalized uncertainty relation
 \be \label{1eq2} \Delta x \Delta p \ge \hbar
\Bigg[1+\alpha^2 l_p^2 \frac{(\Delta p)^2}{\hbar^2}\Bigg] \ee with
$l_p=\sqrt{G \hbar/c^3}$ the Planck length. Here  a parameter
$\alpha=\hbar\sqrt{\beta}/l_p$ is introduced to take into account
the GUP effect. The Planck mass is given by $m_p=\sqrt{\hbar c/G}$.
The above implies a lower bound on the length scale \be \label{2eq2}
\Delta x \ge (\Delta x)_{\rm mim} \approx 2\hbar \sqrt{\beta}=
2\alpha l_p, \ee
 which means that the Planck length plays
the role of a fundamental scale. On the other hand, Eq.
(\ref{1eq2}) implies  the upper bound on  the momentum as \be
\label{2eq2-1} \Delta p \le (\Delta p)_{\rm max} \approx
\frac{2}{\sqrt{\beta}}= \frac{2m_p c}{\alpha}. \ee Furthermore,
 the GUP may be used to derive temperature for the modified Schwarzschild black hole  by identifying  $\Delta p$
 with
 the energy (temperature) of radiated
photons~\cite{MKP}. The momentum uncertainty for radiated photons
can be found to be

\be \label{3eq2} \frac{ \Delta x}{2 \alpha^2 l_p^2}
  \Bigg[1- \sqrt{1-\frac{4\alpha^2 l_p^2}{(\Delta
 x)^2}}\Bigg] \le  \frac{\Delta p}{\hbar} \le \frac{ \Delta x}{2\alpha^2 l_p^2}
  \Bigg[1+ \sqrt{1-\frac{4 \alpha^2 l_p^2}{(\Delta
 x)^2}}\Bigg].
 \ee
 The left inequality implies  small corrections to the
 Heisenberg's uncertainty principle for $\Delta x\gg \alpha l_p$ as $\Delta p \ge
 \hbar/\Delta x +\hbar \alpha^2l_p^2/(\Delta x)^3+\cdots$~\cite{Park}.
On the other hand, the right inequality means  that $\Delta p$
cannot be arbitrarily large in order that the GUP in (\ref{1eq2})
makes sense. For simplicity, we use the Planck units of
$c=\hbar=G=k_B=1$ which imply that $l_p=m_p=1$ and
$\beta=\alpha^2$. Considering the GUP effect on the near-horizon
and $\Delta x=2r_{+}=4M$, the
 relation (\ref{3eq2})  reduces to

\be \label{4eq2} M
  \Bigg[1- \sqrt{1-\frac{\beta }{4M^2}}\Bigg] \le  \frac{\beta \Delta p}{2} \le
  M
  \Bigg[1+ \sqrt{1-\frac{\beta}{4M^2}}\Bigg].
 \ee
Replacing $\beta$ with $2/\omega$, the above leads to a relation
\be \label{analogy} r_-\le \frac{\Delta p}{  \omega} \le r_+.
 \ee
Here we wish to mention that a replacement of  $\beta\to 2/\omega$
was performed  because both sides of Eq. (\ref{4eq2}) have
mathematically the same form as Eq. (\ref{sol}). It seems that Eq.
(\ref{analogy}) indicates a connection between quantum and classical
properties of KS black holes in the deformed HL gravity.

Importantly, it was shown that  based on the GUP~\cite{MV}, quantum
correction to the Bekenstein-Hawking entropy $S_{BH}$  of
Schwarzschild black hole  is given by~\cite{ZZ,WGZ} \be
\label{gupcep} S_{GUP}=S_{BH}+ \pi \alpha^2
\ln[S_{BH}]-\pi^2\alpha^4\frac{1}{2S_{BH}}+\cdots.\ee Considering
the replacement of $\alpha^2=\beta \to 2/\omega$, the above
expression leads to \be \label{gupce} S_{GUP}=S_{BH}+
\frac{2\pi}{\omega}
\ln[S_{BH}]-\frac{4\pi^2}{\omega^2}\frac{1}{2S_{BH}}+\cdots\ee with
$S_{BH}=\pi r_+^2=A/4$.

Promisingly, for $\alpha^2=\beta \to 1/\omega$, Eq. (\ref{gupcep})
recovers Eq. (\ref{entropy}) up to the logarithmic term as
 \be S_{GUP}\simeq \frac{A}{4}+ \frac{\pi}{\omega}
\ln\Big[\frac{A}{4}\Big].\ee

\section{Discussions}
We have found the entropy of Kehagias-Sfetsos black hole in the
deformed HL gravity by using
  the first law of thermodynamics. The presence of logarithmic
  term $\ln[A/4]$ seems to be universal for the HL gravity because it
  appeared
   in topological black hole solutions~\cite{CCO1} and LMP
   solution~\cite{CCO2}.

We would like to mention that  the GUP seems to be a powerful tool
to study quantum gravity effects.  The  GUP with the relation
$\Delta x=2r_+=4M $ of the Schwarzschild radius $r_{+}$ and its ADM
mass $M$ make sense because  quantum gravity effects of GUP is
universal. Thus,  the Schwarzschild black hole was modified  if one
assumes the GUP. It seems  that the GUP  explains a part of quantum
gravity effects but not whole of these.  A relevant relationship of
the black hole entropy-area  based on string theory and loop quantum
gravity is given by~\cite{AAP} \be \label{LQG}
S_{LQG}=\frac{A}{4}+\rho \ln \Big[\frac{A}{4}\Big]+{\cal
O}\Big(\frac{1}{A}\Big),\ee where $\rho$ is a model-dependent
parameter. Therefore, the GUP was widely used to obtain quantum
correction (\ref{LQG})  to the Bekenstein-Hawking entropy of
Schwarzschild black hole~\cite{MV,ZZ,WGZ}.

In this work,  we have attempted  to explain the logarithmic term
$(\pi/\omega)\ln[A/4]$ for the entropy of black hole in the deformed
HL gravity by considering the GUP-inspired entropy to Schwarzschild
black hole. The corresponding quantities  may be $\beta$ in the
generalized commutation relation (\ref{crgup}) and $1/\omega$ of
parameter in the deformed HL gravity (\ref{react}). In the limit of
$\beta \to 0$, we recover the Heisenberg uncertainty relation
without quantum gravity effects, while in the limit of $\omega \to
\infty$, the entropy of Schwarzschild black hole is recovered
without the logarithmic term. This may imply  a close connection
between GUP and HL gravity.

However, we recognize that  the entropy of KS black holes (standard
black hole thermodynamics) was being compared  with that of the
GUP-inspired Schwarzschild black hole (non-standard black hole
thermodynamics). This implies that there exists a sort of
correspondence (duality)  between two systems, but not  that the GUP
is a fundamental property of Ho\v{r}ava-Lifshitz gravity.
Especially, we realize that logarithmic terms in black hole entropy
appeared in many different models. At this stage, thus, the
logarithmic term in the KS black hole entropy does not represent a
definite signal that the GUP is a underlying principle of Ho\v{r}ava
construction.

Consequently, we have shown that the duality in the entropy between
the KS black hole from the HL gravity and GUP-inspired Schwarzschild
black hole is present.

\section*{Acknowledgement}
The author thanks Hyung Won Lee and Yong-Wan Kim for helpful
discussions. This work was in part  supported by Basic Science
Research Program through the National Research Foundation (NRF)  of
Korea funded by the Ministry of Education, Science and Technology
(2009-0086861) and the NRF grant funded by the Korea
government(MEST) through the Center for Quantum Spacetime(CQUeST) of
Sogang University with grant number 2005-0049409.


\begin{thebibliography}{99}

\bibitem{ho1}
P.~Horava,
  Phys.\ Rev.\  D {\bf 79} (2009) 084008
  [arXiv:0901.3775 [hep-th]].


\bibitem{ho2}
  P.~Horava,
  JHEP {\bf 0903} (2009) 020
  [arXiv:0812.4287 [hep-th]].

  \bibitem{LMP}
   H.~Lu, J.~Mei and C.~N.~Pope,
  Phys.\ Rev.\ Lett.\  {\bf 103} (2009) 091301
  [arXiv:0904.1595 [hep-th]].



 \bibitem{CCO1}
  R.~G.~Cai, L.~M.~Cao and N.~Ohta,
  Phys.\ Rev.\  D {\bf 80} (2009) 024003
  [arXiv:0904.3670 [hep-th]].




\bibitem{CLS}
  R.~G.~Cai, Y.~Liu and Y.~W.~Sun,
  JHEP {\bf 0906} (2009) 010
  [arXiv:0904.4104 [hep-th]].



\bibitem{CY}
  E.~O.~Colgain and H.~Yavartanoo,
  JHEP {\bf 0908} (2009) 021
  [arXiv:0904.4357 [hep-th]].

\bibitem{MK}
  Y.~S.~Myung and Y.~W.~Kim,
  arXiv:0905.0179 [hep-th].

\bibitem{CCO2}
  R.~G.~Cai, L.~M.~Cao and N.~Ohta,
  Phys.\ Lett.\  B {\bf 679} (2009) 504
  [arXiv:0905.0751 [hep-th]].


\bibitem{Gho}
  A.~Ghodsi,
  arXiv:0905.0836 [hep-th].

\bibitem{KS}
  A.~Kehagias and K.~Sfetsos,
  Phys.\ Lett.\  B {\bf 678} (2009) 123
  [arXiv:0905.0477 [hep-th]].

\bibitem{Myung}
  Y.~S.~Myung,
  Phys.\ Lett.\  B {\bf 678} (2009) 127
  [arXiv:0905.0957 [hep-th]].


\bibitem{CJ1}
   S.~Chen and J.~Jing,
  arXiv:0905.1409 [gr-qc].
\bibitem{CJ2}
  S.~b.~Chen and J.~l.~Jing,
  Phys.\ Rev.\  D {\bf 80} (2009) 024036
  [arXiv:0905.2055 [gr-qc]].


\bibitem{CW}
 J.~Chen and Y.~Wang,
  arXiv:0905.2786 [gr-qc].

\bibitem{park}
  M.~i.~W.~Park,
  JHEP {\bf 0909} (2009) 123
  [arXiv:0905.4480 [hep-th]].



\bibitem{BGS}

  M.~Botta-Cantcheff, N.~Grandi and M.~Sturla,
  arXiv:0906.0582 [hep-th].



\bibitem{Gho09}
A.~Ghodsi and E. Hatefi
  arXiv:0906.1237 [hep-th].

\bibitem{CL}
  A.~Castillo and A.~Larranaga,
  arXiv:0906.4380 [gr-qc].

\bibitem{PW}
J.~J.~Peng and S.~Q.~Wu,
  arXiv:0906.5121 [hep-th].

  \bibitem{lkme}
  H.~W.~Lee, Y.~W.~Kim and Y.~S.~Myung,
  arXiv:0907.3568 [hep-th].
\bibitem{gup1}
  D.~Amati, M.~Ciafaloni and G.~Veneziano,
  Phys.\ Lett.\  B {\bf 216} (1989) 41;

M.~Maggiore,
  Phys.\ Lett.\  B {\bf 304} (1993) 65
  [arXiv:hep-th/9301067];

  L.~J.~Garay,
  Int.\ J.\ Mod.\ Phys.\  A {\bf 10} (1995) 145
  [arXiv:gr-qc/9403008];

   Y.~J.~Ng and H.~Van Dam,
  Mod.\ Phys.\ Lett.\  A {\bf 9} (1994) 335;

Y.~J.~Ng and H.~Van Dam,
  Mod.\ Phys.\ Lett.\  A {\bf 10} (1995) 2801.




\bibitem{gup2}
A.~Kempf, G.~Mangano and R.~B.~Mann,
  Phys.\ Rev.\  D {\bf 52} (1995) 1108
  [arXiv:hep-th/9412167];


A.~Kempf,
  J.\ Phys.\ A  {\bf 30} (1997) 2093
  [arXiv:hep-th/9604045];

F.~Brau,
  J.\ Phys.\ A  {\bf 32} (1999) 7691
  [arXiv:quant-ph/9905033];

F.~Scardigli,
  Phys.\ Lett.\  B {\bf 452} (1999) 39
  [arXiv:hep-th/9904025];

R.~J.~Adler, P.~Chen and D.~I.~Santiago,
  Gen.\ Rel.\ Grav.\  {\bf 33} (2001) 2101
  [arXiv:gr-qc/0106080].


  \bibitem{CMOK}
  L.~N.~Chang, D.~Minic, N.~Okamura and T.~Takeuchi,
  Phys.\ Rev.\  D {\bf 65} (2002) 125028
  [arXiv:hep-th/0201017].

\bibitem{KLM2}
  Y.~W.~Kim, H.~W.~Lee and Y.~S.~Myung,
  Phys.\ Lett.\  B {\bf 673} (2009) 293
  [arXiv:0811.2279 [gr-qc]].



  \bibitem{myunggup}
  Y.~S.~Myung,
  Phys.\ Lett.\  B {\bf 679} (2009) 491
  [arXiv:0907.5256 [hep-th]].

\bibitem{cos1}
 E. Kiritsis and G. Kofinas,
  arXiv:0904.1334 [hep-th].

\bibitem{muko}
   S.~Mukohyama,
  JCAP {\bf 0909} (2009) 005
  [arXiv:0906.5069 [hep-th]].



\bibitem{DV}
  S.~Das and E.~C.~Vagenas,
  Phys.\ Rev.\ Lett.\  {\bf 101} (2008) 221301
  [arXiv:0810.5333 [hep-th]].


\bibitem{MKP}
  Y.~S.~Myung, Y.~W.~Kim and Y.~J.~Park,
  Phys.\ Lett.\  B {\bf 645} (2007) 393
  [arXiv:gr-qc/0609031].


\bibitem{Park}
  M.~i.~Park,
  Phys.\ Lett.\  B {\bf 659} (2008) 698
  [arXiv:0709.2307 [hep-th]].

  \bibitem{MV}
  A.~J.~M.~Medved and E.~C.~Vagenas,
  Phys.\ Rev.\  D {\bf 70} (2004) 124021
  [arXiv:hep-th/0411022].

\bibitem{ZZ}
  R.~Zhao and S.~L.~Zhang,
  Phys.\ Lett.\  B {\bf 641} (2006) 208.

\bibitem{WGZ}
  F.~J.~Wang, Y.~X.~Gui and Y.~Zhang,
  Gen. Relativ. Gravit., to appear.

\bibitem{AAP}
  G.~Amelino-Camelia, M.~Arzano and A.~Procaccini,
  Phys.\ Rev.\  D {\bf 70} (2004) 107501
  [arXiv:gr-qc/0405084].

\end{thebibliography}
\end{document}